\documentclass[preprintnumbers,amsmath,amssymb,twocolumn]{revtex4}
\usepackage{graphicx}
\usepackage{dcolumn}
\usepackage{bm}
\usepackage{epsfig}

\begin{document}
\title{Phase control of squeezing in fluorescence radiation}
\author{R. Arun}
\affiliation{Department of Physics, School of Basic $\&$ Applied Sciences, Central University of
Tamilnadu, Thiruvarur 610101, Tamilnadu, India.}
\email{rarun@cutn.ac.in}

\begin{abstract}
We study squeezing properties of the fluorescence radiation emitted by a driven $\Lambda$-type
atom in which the metastable lower energy levels are coupled by an additional field. We find that
the relative phase of the applied fields can significantly modify the squeezing characteristics of radiation.
It is shown that the additional field connecting the lower levels in the system can induce spectral squeezing in
a parameter regime for which the squeezing is absent without the additional field. Moreover, the squeezing can be shifted
from inner- to outer-sidebands of the spectrum by simply changing the relative phase. A dressed-state description is presented
to explain these numerical results. The phase control of squeezing in the total variance of quadrature components is also
examined. We show that the squeezing in total variance attains its maximal value when the system reduces to an effective two-level
system.
\end{abstract}

\maketitle

\newpage
\section{Introduction}

Squeezing of the radiation emitted in resonance fluorescence of driven atoms has been extensively investigated
over the last couple of decades \cite{review}. Squeezed states of light have a reduced variance in quadrature
components of the electric field below its shot-noise limit \cite{mandel}. Theoretical studies considered either
the total variance of phase quadratures or the squeezing spectrum of fluorescence radiation to demonstrate
squeezing \cite{zoller,collet,swain,oliveira,vogel,vee,ladder,dalton,vogel2,gao}.  Walls and Z\"{o}ller first
predicted total variance squeezing in the fluorescent light of driven two-level systems \cite{zoller}. The calculations
on the squeezing spectrum demonstrated single- and two-mode squeezing in the weak- and strong-excitation regimes
\cite{collet,swain}. The studies on squeezing have been extended to three-level systems in $\Lambda$ \cite{oliveira,vogel},
V \cite{vee} and ladder \cite{ladder} configurations. Dalton {\it et al} \cite{dalton} examined the role of atomic coherence
on the squeezing in fluorescence from three-level systems. It was shown that maximal squeezing is obtained when the atomic
system evolves into a pure state \cite{dalton}. Recently, Gr\"{u}nwald and Vogel \cite{vogel2} have proposed an ingenious scheme
using cavity-assisted purification to achieve near-maximal squeezing in fluorescence. A detailed study by Gao {\it et al}
\cite{gao} has shown that the squeezing spectrum of three-level atoms may exhibit ultranarrow peaks.

In all these publications, the squeezing properties of the fluorescence radiation are obtained independent of the phases of
applied lasers. Recently, much attention has been paid on the control of medium properties by the phases of
applied fields \cite{vic,trap,hemmer,ionize,probes,eit1,eit2,scully,sun,quench,dark}. One way to achieve phase control is by vacuum
induced coherences which arise due to the atomic transitions coupled by same vacuum modes \cite{vic}. An alternative way is to
use a closed-loop scheme of transitions in atoms \cite{trap,hemmer,ionize,probes,eit1,eit2,scully,sun,quench,dark}. In this scheme,
phase-dependent behavior has been reported in both the dynamics and steady-state properties of driven systems \cite{trap}.
Many interesting effects have been studied on the phase control of population dynamics \cite{trap}, photoionization \cite{ionize},
preparation of microwave-spin dressed states \cite{hemmer}, quantum interferences in probe absorption \cite{probes},
electromagnetically induced transparency (EIT) \cite{eit1,eit2,scully}, fast and slow light propagation \cite{sun}, fluorescence
quenching and line narrowing \cite{quench}. In $\Lambda$-type systems controlled by a microwave field coupling the ground states,
it has been shown experimentally that the trapping state evolves into a microwave-spin dressed state for an appropriate choice of
laser phases \cite{hemmer}. Further, phase-dependent effects on probe light absorption \cite{probes}, EIT in double-$\Lambda$ system
\cite{eit1}, splitting of EIT transparencies \cite{eit2}, and probe transmission in EIT \cite{scully} have also been experimentally
demonstrated. A recent theoretical study \cite{dark} has discussed the rigorous dark state conditions required to establish EIT in a
$\Lambda$-type system with closed-loop transitions.

\begin{figure}[b]
    \includegraphics[width=6cm]{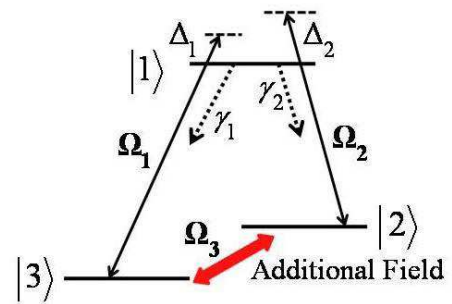}
\caption{The level scheme of a closed $\Lambda$-type three-level atom.}
\end{figure}
In this paper, we consider a closed $\Lambda$-type system interacting with two coherent fields (as shown in Fig. 1). It is assumed
that the excited atomic state decays spontaneously to the ground states which are metastable. We further assume that an additional
field couples the metastable ground states as in earlier publications \cite{hemmer,scully,quench,dark}. This additional field could be a
microwave, infrared or a rf field depending upon the level spacings. The role of the additional field and its phase control was
investigated in the fluorescence spectrum of this system in Ref. \cite{quench}. In the present work, we study the squeezing
properties of the fluorescence field and examine how the relative phase of applied fields can modify squeezing aspects.
The outline of the paper is as follows: Section II establishes the model and basic dynamical equations which govern the
atom-field interaction including decay processes. In Section III, we present the numerical results and study the effects of additional
field on the squeezing spectrum. Section IV is devoted to analyze the squeezing in total variance optimized with respect to parameters
of the system. Finally, the main results are summarized in Section V.

\section{Model system and density matrix equations}
We consider a three-level atom of the $\Lambda$ configuration driven by three fields as shown in figure 1. The excited state $|1\rangle$
is driven to the ground states $|3\rangle$ and $|2\rangle$ by two coherent fields (frequencies $\omega_1,\omega_2$ and phases
$\phi_1,\phi_2$) of Rabi frequencies $\Omega_1$ and $\Omega_2$, respectively. The ground states $|2\rangle$ and $|3\rangle$,
being metastable states, are coupled by an additional field (frequency $\omega_3$ and phase $\phi_3$) of Rabi frequency $\Omega_3$.
We assume that the atom decays by spontaneous emission along the channels $|1\rangle \rightarrow |3\rangle$ and
$|1\rangle \rightarrow |2\rangle$ with rates $2 \gamma_1$ and $2 \gamma_2$, respectively. The atomic dynamics is studied in an
appropriate rotating frame and by changing the phases of basis states as $|1\rangle \rightarrow e^{i \phi_1} |1\rangle$ and
$|2\rangle \rightarrow e^{i (\phi_1 - \phi_2)} |2\rangle$. In the rotating frame and new basis states, the Hamiltonian is given in
the dipole approximation as
\begin{flalign}
H =& -\hbar \Delta_1 |1\rangle \langle 1| - \hbar (\Delta_1 - \Delta_2) |2\rangle \langle 2| && \nonumber \\
& -\hbar (\Omega_1 |1\rangle \langle 3| + \Omega_2 |1\rangle \langle 2| + \Omega_3 e^{i(\Delta_4 t + \Phi)} |2\rangle \langle 3|
+ \hbox{h.c.}). &&\label{hami}
\end{flalign}
Here, $\Delta_1$ ($\Delta_2$) denotes the detuning of the field driving the transition $|1\rangle \leftrightarrow |3\rangle$
($|1\rangle \leftrightarrow |2\rangle$). Similarly, $\Delta_3$ corresponds to the detuning of the additional field coupling the
transitions $|2\rangle \leftrightarrow |3\rangle$. The relative detuning $\Delta_4 = \Delta_1 - \Delta_2 - \Delta_3 = \omega_1 - \omega_2
- \omega_3$ gives the frequency difference and the relative phase $\Phi = \phi_1-\phi_2-\phi_3$ represents the phase difference of the
applied fields.

We use the master equation framework to include spontaneous emission in the atomic dynamics. With the inclusion of decay terms,
the density matrix elements obey the following time-dependent equations
\begin{equation}
\dot{\rho}_{11} = -2(\gamma_1 + \gamma_2) \rho_{11} + i \Omega_1 \rho_{31} + i \Omega_2
\rho_{21} - i \Omega_1 \rho_{13} - i \Omega_2 \rho_{12},
\label{rho1}
\end{equation}
\begin{flalign}
\dot{\rho}_{22} = &~ 2 \gamma_2 \rho_{11} + i \Omega_2 \rho_{12} + i \Omega_3 e^{i(\Delta_4 t + \Phi)} \rho_{32} - i \Omega_2 \rho_{21} &&
\nonumber \\
& - i \Omega_3 e^{-i(\Delta_4 t + \Phi)} \rho_{23},&&
\end{flalign}
\begin{flalign}
\dot{\rho}_{12} = &-(\gamma_1 + \gamma_2 - i \Delta_2) \rho_{12} + i \Omega_2 (\rho_{22} - \rho_{11}) + i \Omega_1 \rho_{32} &&
\nonumber \\
& - i \Omega_3 e^{-i(\Delta_4 t + \Phi)} \rho_{13},&&
\end{flalign}
\begin{flalign}
\dot{\rho}_{13} = &-(\gamma_1 + \gamma_2 - i \Delta_1) \rho_{13} + i \Omega_1 (\rho_{33} - \rho_{11}) + i \Omega_2 \rho_{23} &&
\nonumber \\
&- i \Omega_3 e^{i(\Delta_4 t + \Phi)} \rho_{12},&&
\end{flalign}
\begin{flalign}
\dot{\rho}_{23} = &~ i (\Delta_1 - \Delta_2) \rho_{23} + i \Omega_3 e^{i(\Delta_4 t + \Phi)} (\rho_{33} - \rho_{22}) - i \Omega_1 \rho_{21}
&& \nonumber \\
& + i \Omega_2 \rho_{13},&& \label{rho5}
\end{flalign}
with $\rho_{ji} = \rho_{ij}^{*}$ and $\rho_{11} + \rho_{22} + \rho_{33} = 1$. It is seen from equations (\ref{rho1})
- (\ref{rho5}) that the exponential phase terms $e^{i \Phi}$ are always accompanied by the additional field Rabi
frequency $\Omega_3$. This shows that the atomic dynamics becomes dependent on the relative phase $\Phi$ only when the
additional field is applied on the system.

In what follows we assume the frequencies of the applied fields to satisfy the condition $\omega_1 = \omega_2
+ \omega_3$ which implies the relative detuning $\Delta_4$ to be zero. The explicit time dependence in equations
(\ref{rho1})-(\ref{rho5}) is then removed and the equations can be easily solved in steady state. For convenience
in the calculation of steady-state properties, we rewrite the density matrix equations (\ref{rho1})-(\ref{rho5})
in a more compact matrix form by the definition
\begin{equation}
\hat{\Psi} = \left(\rho_{11},\rho_{22},\rho_{12},\rho_{21},\rho_{13},\rho_{31},\rho_{23},\rho_{32}\right)^{T}.
\label{psidef}
\end{equation}
Substituting equation (\ref{psidef}) into equations (\ref{rho1})-(\ref{rho5}) with $\Delta_4 = 0$, the
matrix equation for the variables $\hat{\Psi}_j(t)$ obeys
\begin{equation}
\frac{d}{dt} \hat{\Psi} = \hat{L} \hat{\Psi} + \hat{I}, \label{matrix}
\end{equation}
where $\hat{\Psi}_j$ is the $j$-th component of the column vector $\hat{\Psi}$ and the
inhomogeneous term $\hat{I}$ is also a column vector with non-zero components
\begin{flalign}
&\hat{I}_5 = i \Omega_1, \hat{I}_6 = -i \Omega_1, \hat{I}_7 = i \Omega_3 e^{i\Phi},
\hat{I}_8 = -i \Omega_3 e^{-i\Phi}.&&
\end{flalign}
In equation (\ref{matrix}), $\hat{L}$ is a 8$\times$8 matrix whose elements are
time independent and can be found explicitly from equations (\ref{rho1})-(\ref{rho5}). The steady-state
values of the density matrix elements can be obtained by setting the time derivative equal to
zero in equation (\ref{matrix}):
\begin{equation}
\hat{\Psi}(\infty) = - \hat{L}^{-1} \hat{I}. \label{steady}
\end{equation}

\section{calculation of the squeezing spectrum}
Since the atom is driven by two coherent fields, each field produces its own fluorescence field from
the system. However, the fluorescence fields generated by the $|1\rangle \Leftrightarrow |3\rangle$ and
$|1\rangle \Leftrightarrow |2\rangle$ transitions in the atom will have no correlations because the
frequencies $\omega_1$ and $\omega_2$ of the applied fields driving the transitions are quite different.
We consider squeezing in the fluorescent light exclusively emitted by the $|1\rangle \Leftrightarrow |3\rangle$
transitions in the atom. Assuming that the detection of fluorescence field is in a direction perpendicular to
the atomic dipole moment, the positive and negative frequency parts of the electric field operator in the
radiation zone can be written as
\begin{eqnarray}
\vec{E}^{(+)}(t) &=& f(r) \vec{\mu}_{13} A_{31}(\hat{t})
\exp[-i(\omega_1\hat{t} + \phi_1)], \label{electric} \\
\vec{E}^{(-)}(t) &=& [\vec{E}^{(+)}(t)]^{\dagger}, \nonumber
\end{eqnarray}
where $\hat{t} = t - r/c$, $f(r) = \omega_{13}^2/c^2r$, $r$ is the distance of the detector from the atom, and
the operators $A_{mn} = |m\rangle\langle n|$ represent the transition operators for $m \neq n$ and atomic
population operators for $m=n$.
To calculate the squeezing spectrum, we introduce slowly varying quadrature components with phase $(\theta)$ as
\begin{equation}
\vec{E}(\theta,t) = \vec{E}^{(+)}(t) e^{i(\omega_1 t+\theta)} + \vec{E}^{(-)}(t) e^{-i(\omega_1 t+\theta)}.
\label{quad}
\end{equation}
The squeezing spectrum is defined by the Fourier transformation of the normal and time-ordered
correlation of the quadrature component $\vec{E}(\theta,t)$:
\begin{equation}
S(\omega,\theta) = \frac{1}{2\pi} \int_{-\infty}^{\infty} \hat{T}
\langle:\vec{E}(\theta,t),\vec{E}(\theta,t+\tau):\rangle e^{i\omega\tau}d\tau, \label{spect}
\end{equation}
where $\langle\vec{A},\vec{B}\rangle = \langle\vec{A}.\vec{B}\rangle - \langle\vec{A}\rangle .
\langle\vec{B}\rangle$ and $\hat{T}$ represents the time ordering operator.

In the steady-state limit $(t \rightarrow \infty)$, the correlation function appearing in equation
(\ref{spect}) can be easily obtained using the quantum regression theorem and the density matrix
equations (\ref{matrix}). For the purpose of calculations, we introduce column vectors of two-time averages
\begin{flalign}
&\hat{U}^{mn}(t,\tau) = && \nonumber \\
 &\left[\langle \Delta A_{11}(t+\tau) \Delta A_{mn}(t)\rangle, \langle \Delta A_{22}(t+\tau)\Delta A_{mn}(t)\rangle,
\right. &&\nonumber \\
& ~\langle \Delta A_{21}(t+\tau) \Delta A_{mn}(t)\rangle, \langle \Delta A_{12}(t+\tau) \Delta A_{mn}(t)\rangle, &&\nonumber\\
& ~\langle \Delta A_{31}(t+\tau) \Delta A_{mn}(t)\rangle, \langle \Delta A_{13}(t+\tau) \Delta A_{mn}(t)\rangle, &&\nonumber \\
& \left. \langle \Delta A_{32}(t+\tau) \Delta A_{mn}(t)\rangle, \langle \Delta A_{23}(t+\tau) \Delta A_{mn}(t)\rangle \right]^{T},
&& \nonumber \\
&\hskip 2.1in m,n = 1,2,3.&& \label{corre}
\end{flalign}
Here, $\Delta A_{mn}(t) = A_{mn}(t) - \langle A_{mn}(\infty) \rangle$ are the deviations of the
atomic operators from its steady-state values (\ref{steady}). Now, applying the quantum regression theorem and time ordering
of operators in equation (\ref{spect}) as explained in \cite{sref}, the squeezing spectrum can be obtained as
\begin{flalign}
S(\omega,\theta) = & \frac{{|\vec{\mu}_{13}|}^2 f(r)^2}{\pi} \hbox{Re}\Bigg\{\sum_{k = 1}^{8}
\lim_{t \rightarrow \infty} \left[ \hat{M}_{5,k} \hat{U}^{31}_k(t,0) \right.  &&\nonumber \\
&~~\left. \times e^{2i(\theta - \phi_1 + \omega_1 r/c)} + \hat{M}_{6,k} \hat{U}^{31}_k(t,0) \right]\Bigg\},&&
\label{mains}
\end{flalign}
where $\hat{M}_{j,k}$ denotes the $(j,k)$ element of the matrix $\hat{M} = [(i \omega - \hat{L})^{-1} + (-i \omega - \hat{L})^{-1}]$.

\begin{figure}
   \includegraphics[width=8cm]{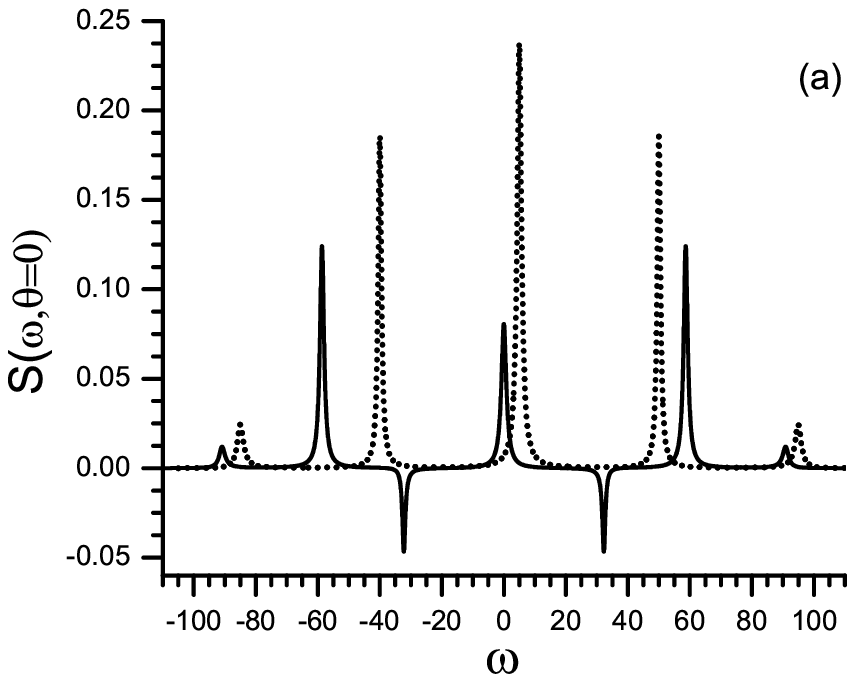} \\
   \includegraphics[width=8cm]{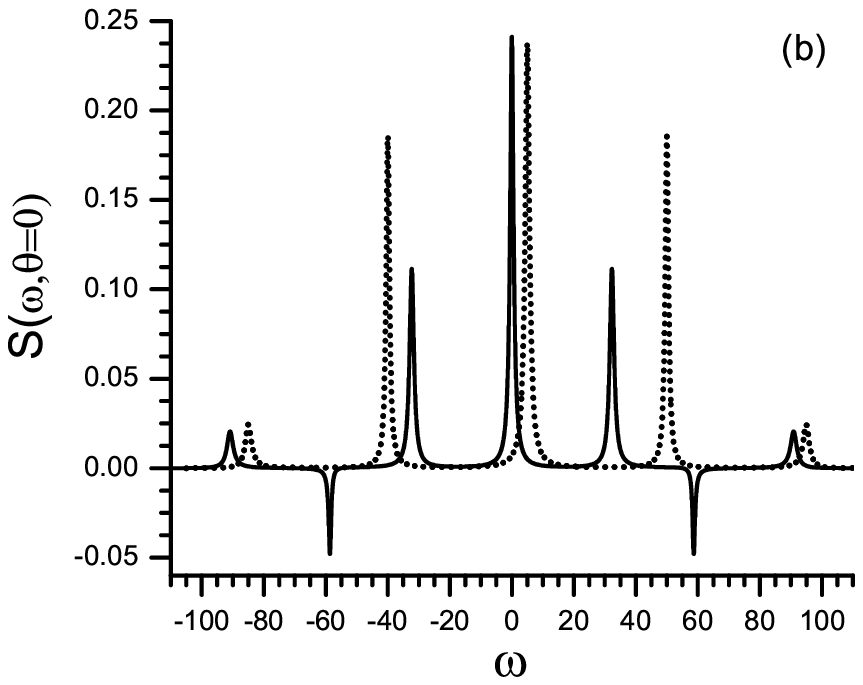}
   \vspace*{-0.3cm}
   \caption{Squeezing spectrum $S(\omega,\theta)$ as a function of $\omega$ for the parameters $\theta = 0$, $\gamma_1 = 0.1$, $\gamma_2 = 1$,
   $\Delta_1 = -\Delta_2 = 15$, $\Omega_1 = \Omega_2 = 30$, and $\Phi = 0$ (a) and $\Phi = \pi$ (b). The solid (dotted) curves are for
   $\Omega_3 = 10$ $(\Omega_3 = 0)$. For clarity, the dotted curve has been displaced by 5 units along the $\omega$-axis. Actual values of the
   dotted curve are 0.5 times that shown. }
\end{figure}
We now proceed to present the numerical results of the squeezing spectrum and its interpretation. From the definition of squeezing
\cite{mandel}, a fluorescence light exhibits spectral squeezing in a selected quadrature component $(\theta)$ if the squeezing
spectrum is negative, $S(\omega,\theta) < 0$, at a certain frequency $\omega$. To demonstrate this squeezing in spectral components,
we analyze numerically the spectrum calculated using equation (\ref{mains}) for a special parameter choice $\Omega_1 = \Omega_2$ and
$\Delta_1 = -\Delta_2$. In the numerical calculation, all the frequency parameters such as decay rates, Rabi frequencies and detunings
are scaled in units of $\gamma_2$. We also assume $e^{2i(-\phi_1 + \omega_1 r/c)} = 1$ and scale the spectrum in units of
$\mu_{13}^2 f(r)^2/(\pi \gamma_2)$. In figure 2 the numerical results \cite{note1} are presented for the in-phase quadrature $(\theta = 0)$
with two different values of the relative phase ($\Phi = 0, \pi$). The graphs show that the squeezing is absent in the spectrum
[see dashed curves in figure 2] when the additional field is not applied on the system $(\Omega_3 = 0)$ \cite{note2}. Note that the
squeezing spectrum is independent of the relative phase $\Phi$ of the applied fields without the additional field $(\Omega_3 = 0)$ as
expected. Interesting features appear in the spectrum only when the additional field connects the lower metastable levels in the system.
As seen in figure 2, the squeezing is induced in the spectrum depending on the relative phase $(\Phi)$ for $\Omega_3 \neq 0$. The spectral
squeezing is shifted from inner- to outer-sidebands of the spectrum as the relative phase is changed from $\Phi = 0$ to $\Phi = \pi$
[compare solid curves in figure 2].

To explain the origin of the new features in the squeezing spectrum, we go to the dressed-state description of the atom-field interaction.
The dressed states $|\Phi_i\rangle (i = \alpha, \beta, \kappa)$ defined as eigenstates $(H |\Phi\rangle = \hbar \lambda |\Phi\rangle)$ of the
Hamiltonian (\ref{hami}) can be expanded in terms of the bare atomic states as
\begin{equation}
|\Phi_i\rangle = a_{1i} |1\rangle + a_{2i} |2\rangle + a_{3i} |3\rangle ,
\end{equation}
where the expansion coefficients are explicitly given by
\begin{eqnarray}
a_{1i} &=& N[\lambda_i \Omega_2 - \Omega_1 \Omega_3 \exp(-i \Phi)],\nonumber \\
a_{2i} &=& N[\Omega_1^2 - \lambda_i(\Delta_1 + \lambda_i)],\nonumber \\
a_{3i} &=& N[(\Delta_1 + \lambda_i) \Omega_3 \exp(-i \Phi) - \Omega_1 \Omega_2].\label{ecoeff}
\end{eqnarray}
Here, the overall constant factor N is appropriately chosen to satisfy the normalization condition $|a_{1i}|^2 + |a_{2i}|^2 + |a_{3i}|^2 = 1$
\cite{note3}. The eigenvalues $\lambda_i (i = \alpha, \beta, \kappa)$ can be obtained numerically by solving the characteristic equation of the Hamiltonian (\ref{hami}) in the basis of bare atomic states. In order to understand the spectral features, one has to consider the allowed transitions
$|\Phi_i\rangle \leftrightarrow |\Phi_j\rangle (i,j = \alpha, \beta, \kappa)$ between the dressed states including decay processes. The peaks
in the squeezing spectrum occur at the frequencies $\omega_{ij} = \lambda_i - \lambda_j$ due to the dressed-state transitions
$|\Phi_i\rangle \leftrightarrow |\Phi_j\rangle$.

In the high field limit $(\Omega_1,\Omega_2,\Omega_3 \gg \gamma_1, \gamma_2)$, the squeezing
spectrum (\ref{mains}) can be worked out in the dressed-state basis. The contribution to the spectrum by the dressed-state transitions
$|\Phi_{\alpha}\rangle \leftrightarrow |\Phi_{\beta}\rangle$ can be given as
\begin{flalign}
S(\omega_{\pm},0) = &\Gamma_{\alpha\beta}\frac{(a_{1\alpha} a_{3\beta} + a_{3\alpha} a_{1\beta})}{\Gamma_{\alpha\beta}^2 +
{(\omega \mp \omega_{\alpha \beta})}^2} && \nonumber \\
&~~~\times \left[a_{1\alpha} a_{3\beta} \rho_{\alpha \alpha} + a_{3\alpha} a_{1\beta} \rho_{\beta \beta}\right],&&
\label{dress}
\end{flalign}
where the subindex + (-) stands for the positive $(\omega > 0)$ [negative $(\omega < 0)$] part of the spectrum and $\rho_{\alpha \alpha}
(\rho_{\beta \beta})$ represents the population of the dressed state $|\Phi_{\alpha}\rangle (|\Phi_{\beta}\rangle)$. Equation (\ref{dress})
shows that the spectrum  is a pair of Lorentzian curves centered at $\omega = \pm \omega_{\alpha \beta}$ with its width proportional to the decay
rate $\Gamma_{\alpha\beta}$ of dressed-state coherence. The explicit form of the decay rate $\Gamma_{\alpha\beta}$ is given in Appendix A. By using
the numerical values of the expansion coefficients (\ref{ecoeff}), the formula (\ref{dress}) reproduces well the squeezing peaks shown in
figure 2. Specifically in the presence of additional field $(\Omega_3 \neq 0)$, the numerical values of eigenvalues (in units of $\gamma_2$)
for the parameters of figure 2 are $\lambda_{\alpha} = 26.07, \lambda_{\beta} = -6.21, \lambda_{\kappa} = -64.86~(\Phi = 0)$ and $\lambda_{\alpha} =
-56.07, \lambda_{\beta} = -23.79, \lambda_{\kappa} = 34.86~(\Phi = \pi)$. Thus, the inner- and outer- sidebands in the spectrum (solid curves in figure 2)
can be seen as arising from the dressed-state transitions $|\Phi_{\alpha}\rangle \leftrightarrow |\Phi_{\beta}\rangle$ and $|\Phi_{\beta}\rangle
\leftrightarrow |\Phi_{\kappa}\rangle$, respectively. In this case, only a single dressed-state transitions  $|\Phi_i\rangle \leftrightarrow
|\Phi_j\rangle$ contributes to each of the peaks in the squeezing spectrum. However, the situation differs significantly when there is no
additional field acting on the system. For $\Omega_3 = 0$, the numerical values of eigenvalues for the case shown as dashed curves in figure 2
are $\lambda_{\alpha} = -60, \lambda_{\beta} = -15$, and $\lambda_{\kappa} = 30$. It is seen that the outer-sidebands originate from the transitions $|\Phi_{\alpha}\rangle \leftrightarrow |\Phi_{\kappa}\rangle$ of the dressed states. In the case of inner-sidebands peaked at $\omega = \pm 45$,
both the dressed-state transitions $|\Phi_{\alpha}\rangle \leftrightarrow |\Phi_{\beta}\rangle$ and $|\Phi_{\beta}\rangle \leftrightarrow
|\Phi_{\kappa}\rangle$ contribute to the spectrum. The inner-sideband spectrum is then a sum of two different Lorentzians of the form (\ref{dress})
with different widths. The net effect is that the spectral squeezing is absent in the fluorescence field.

\section{squeezing in total variance}
A light field $\vec{E}(\theta,t)$ in a selected quadrature $(\theta)$ is said to be squeezed if the variance $\left<[\Delta\vec{E}(\theta,t)]^2\right>$
is below its value in vacuum state. An equivalent criterion for squeezing \cite{mandel} is that the normal ordered variance of the field
$\left<:[\Delta\vec{E}(\theta,t)]^2:\right>$ is negative. Using the expression (\ref{quad}) of the quadrature component, the normal ordered variance
is defined by
\begin{widetext}
\begin{equation}
\left<:[\Delta\vec{E}(\theta,t)]^2:\right> = \left<(\Delta\vec{E}^{(+)})^2\right> \exp[2i(\omega_1 t + \theta)]
+ \left<(\Delta\vec{E}^{(-)})^2\right> \exp[-2i(\omega_1 t + \theta)] + 2 \left<\Delta\vec{E}^{(-)}\cdot \Delta\vec{E}^{(+)}\right>,
\label{nvar}
\end{equation}
where $\Delta\vec{E}^{(\pm)} = \vec{E}^{(\pm)} - \left<\vec{E}^{(\pm)}\right>$. For the case of a single atom emitting fluorescence radiation,
the operators $\vec{E}^{(\pm)}$ in equation (\ref{nvar}) are replaced by the source-field operators (\ref{electric}). With this substitution, the
normal order variance (\ref{nvar}) reduces further to \cite{vogel}
\begin{equation}
\left<:[\Delta\vec{E}(\theta,t)]^2:\right> = {|\vec{\mu}_{13}|}^2 f(r)^2 \left[2 \rho_{11}(\hat{t}) - 4 |\rho_{13}(\hat{t})|^{2} \cos^2(\theta -
\phi_1 - \phi_{31} + \omega_1 r/c)\right].
\label{vars}
\end{equation}
\end{widetext}
Here, $\phi_{31}$ denotes the phase of the matrix element $\rho_{31} = |\rho_{31}| \exp(i \phi_{31})$. Obviously, if the $\cos^2$ term in
the above equation is unity, the field variance will be minimum. Considering the steady-state limit $(t\rightarrow\infty)$ in equation
(\ref{vars}), the phase-optimized (minimal) normal ordered field variance (denoted as squeezing parameter F) is now given by
\begin{equation}
F \equiv \frac{\left<:[\Delta\vec{E}(\theta,t)]^2:\right>}{{|\vec{\mu}_{13}|}^2 f(r)^2} = 2 \rho_{11} - 4 |\rho_{13}|^{2}, \label{mvar}
\end{equation}
where $\rho_{11}$ and $\rho_{13}$ refer to the steady-state values (\ref{steady}) of the population and coherence, respectively.
\begin{figure}
    \includegraphics[width=8cm]{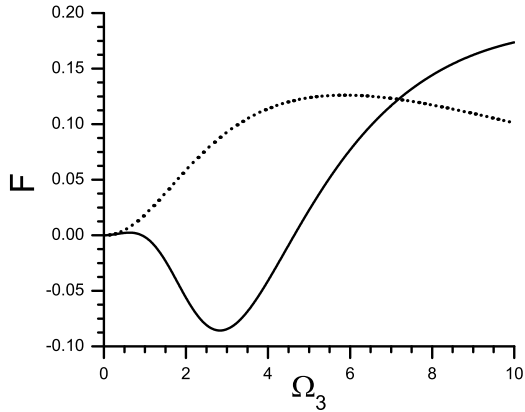}
    \vspace*{-0.3cm}
\caption{The squeezing parameter $F$ versus the Rabi frequency $\Omega_3$ for the parameters $\gamma_1 = 20$, $\gamma_2 = 1$, $\Delta_1 = \Delta_2
= 0$, $\Omega_1 = \Omega_2 = 8$, and $\Phi = -\pi/2$ (solid curve) and $\Phi = \pi/2$ (dotted curve).}
\end{figure}

The equation (\ref{mvar}) can be used to study the squeezing in total field variance optimized with respect to the quadrature phase $(\theta)$.
In the absence of the additional field $(\Omega_3 = 0)$, the driven $\Lambda$ system is known to exhibit fluorescence squeezing when one of
the two transitions in the atom is detected \cite{vogel}. It was shown \cite{vogel} that the squeezing $(F<0)$ occurs in the fluorescence
field only if the decay rate of the detected transition is greater than that of the neighboring transition, i.e., $\gamma_1 > \gamma_2$.
Our numerical analysis shows that this is true even in the present case of the atom subject to an additional field coupling the ground
states. Therefore, we focus only on the fast-decaying transitions in the atom $(\gamma_1 > \gamma_2)$. For simplicity, we consider the
case of resonant light fields $(\Delta_1 = \Delta_2 = 0)$ for a parameter choice $\Omega_1 = \Omega_2 = \Omega$. In this case, the
squeezing parameter F can be obtained in an analytical form to be
\begin{equation}
F = \frac{4 \Omega^2 \Omega_3^2 \sin^2(\Phi)}{M^2} [G - H], \label{mform}
\end{equation}
where
\begin{flalign}
&G = \Omega^4 + \Omega_3^4 + \Omega_3^2[(\gamma_1+ \gamma_2)^2+\Omega^2], &&\\
&H = 2 \Omega_3 \sin(\Phi) [(\gamma_1+ \gamma_2) \Omega_3^2 - 2 \gamma_1 \Omega^2] + 3 \Omega^2 \Omega_3^2 \cos(2\Phi), &&
\end{flalign}
and
\begin{flalign}
M = &~ 2 \Omega^4-\Omega^2 \Omega_3^2 + 2 \Omega_3^2 [(\gamma_1 + \gamma_2)^2 + \Omega_3^2] &&\\ \nonumber
& + \Omega^2 \Omega_3 [2 (\gamma_1 - \gamma_2) \sin(\Phi) - 3 \Omega_3 \cos(2\Phi)].&&
\end{flalign}

\begin{figure}
    \includegraphics[width=8cm]{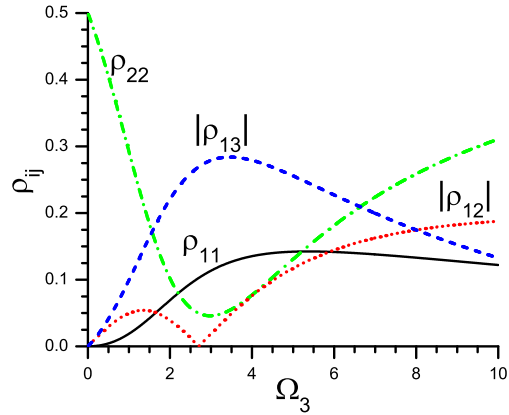}
    \vspace*{-0.3cm}
\caption{(Color online) Steady-state values of the populations and coherences versus the Rabi frequency $\Omega_3$ for $\Phi = -\pi/2$ : $\rho_{11}$
(solid curve), $\rho_{22}$ (dot-dashed curve), $|\rho_{12}|$ (dotted curve), and $|\rho_{13}|$ (dashed curve). The other parameters for the calculation
are the same as in figure 3.}
\end{figure}
Figure 3 displays the squeezing parameter $F$ calculated using (\ref{mform}) as a function of the Rabi frequency $\Omega_3$ for
the relative phases $\Phi = -\pi/2$ and $\Phi = \pi/2$ \cite{note1}. From the graph it is seen that the squeezing parameter
goes negative (positive) for $\Phi = -\pi/2$ $(\Phi= \pi/2)$ as the Rabi frequency $\Omega_3$ is increased. This shows clearly that the
squeezing $(F < 0)$ is induced in the fluorescence field for $\Phi = -\pi/2$ due to the additional field $(\Omega_3 \neq 0)$ acting on the
system. Note further that the squeezing is absent $(F = 0)$ irrespective of the relative phase for $\Omega_3 = 0$. This is due to the well-known
coherent population trapping effect which inhibits the atom from fluorescing. An important result in figure 3 is that the squeezing parameter
exhibits a minimum value (maximum squeezing) as a function of $\Omega_3$ for $\Phi = -\pi/2$ (solid curve). We have found numerically that this
behavior is generally present even in the case of nonzero detunings $(\Delta_1 \neq 0, \Delta_2 \neq 0$) of the driving fields. In order to understand
this result, we plot in figure 4 the steady-state values (\ref{steady}) of the populations and coherences versus $\Omega_3$ for the same parameters
of figure 3. On comparing the figures 3 and 4, it is observed that the population $\rho_{22}$ and the coherence $|\rho_{12}|$ along the
$|1\rangle \leftrightarrow |2\rangle$ transitions are approximately zero near the value of $\Omega_3$ for maximal squeezing. This implies that the
atomic system behaves much like a two-level system along the transitions $|1\rangle \leftrightarrow |3\rangle$. The population in state $|2\rangle$
is forced to return to state $|3\rangle$ rapidly by the action of the additional field. This feature may persist even if one replaces the additional
field with a relaxation between the ground atomic states \cite{oliveira}.

\begin{figure}
    \includegraphics[width=8cm]{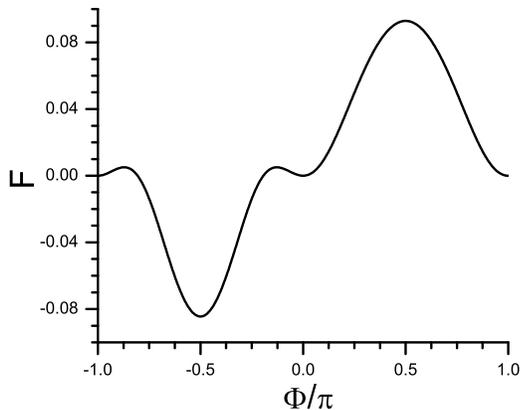}
    \vspace*{-0.3cm}
\caption{The squeezing parameter $F$ as a function of the relative phase $\Phi$ with $\gamma_1 = 20$, $\gamma_2 = 1$, $\Delta_1 = \Delta_2
= 0$, $\Omega_1 = \Omega_2 = 8$, and $\Omega_3 = 3$.}
\end{figure}
It should be emphasized that squeezing in the resonance fluorescence of $\Lambda$ systems has been already investigated
\cite{oliveira,vogel}. In these publications, the fluorescence squeezing is shown to exist independent of the phases of applied lasers.
However, the present paper has studied the dependence of squeezing in the fluorescence field on the relative phase of the applied fields
(see figure 3). To demonstrate further the phase control of squeezing, the squeezing parameter $(F)$ is plotted as a function of the relative phase
$(\Phi)$ in figure 5. The figure clearly indicates that the squeezing is present (absent) for $\Phi < 0$ $(\Phi > 0)$. Thus, one finds that the
presence of an additional field induces phase-dependent squeezing of the radiation. The maximal squeezing occurs only, if $\Phi = -\pi/2$, in which
case the system reduces to an effective two-level system as discussed above. Finally, we note that the squeezing in spectral components does not
guarantee the occurrence of squeezing in total variance \cite{mandel}. The spectral squeezing (see figure 2) may appear even if the detected transition
decays slowly relative to the other transition $(\gamma_1 < \gamma_2)$. However, the squeezing in total variance exists only in the fluorescence field
from the fast-decaying atomic transitions $(\gamma_1 > \gamma_2)$.

\section{conclusions}
In this work we present a theoretical investigation of squeezing properties of the fluorescence radiation from a $\Lambda$ system driven by two coherent
fields and an additional field. In particular, we consider the case when each coherent field drives one transition and the additional field couples the
metastable ground levels of the system. It is seen that the presence of the additional field induces spectral squeezing in the fluorescence. In contrast
to the results of previous studies, the squeezing characteristics now exhibit a strong dependence on the relative phase of the applied fields.
We show that the squeezing peaks in the spectrum can be shifted between the inner- and outer-sidebands just by changing the relative phase.
A description based on dressed states has been given to explain these features. Further, we also investigate the influence of the additional field on
squeezing in the total variance of the fluorescence field. The results show that the squeezing in total variance becomes phase-dependent and attains
a maximal value for a particular relative phase. Moreover, when the maximal squeezing occurs, the system behaves like an effective two-level
system with only the one-photon coherence contributing to the dynamics.

\appendix*
\section{A}
In the secular approximation, the coherence term $\rho_{\alpha \beta}(t)$ in the dressed-state basis obeys
\begin{equation}
\frac{d\rho_{\alpha \beta}}{dt} = -(\Gamma_{\alpha \beta} + i \omega_{\alpha \beta}) \rho_{\alpha \beta,}
\end{equation}
with $\omega_{\alpha \beta} = \lambda_{\alpha} - \lambda_{\beta}$. The decay rate $\Gamma_{\alpha \beta}$ is given by \cite{note3}
\begin{equation}
\Gamma_{\alpha \beta} = \Gamma_1 \gamma_1 + \Gamma_2 \gamma_2,
\end{equation}
where
\begin{eqnarray}
\Gamma_1 &=& a_{1\alpha}^2 + a_{1\beta}^2 - 2 a_{1\alpha} a_{1\beta} a_{3\alpha} a_{3\beta},\nonumber \\
\Gamma_2 &=& a_{1\alpha}^2 + a_{1\beta}^2 - 2 a_{1\alpha} a_{1\beta} a_{2\alpha} a_{2\beta}. \nonumber
\end{eqnarray}


\begin{thebibliography}{99}
\bibitem{review} Special issue on squeezed light, edited by
R. Loudon and P.L. Knight 1987 {\it J. Mod. Opt.} {\bf 34} 6/7; Special issue on squeezed
light, edited by H.J. Kimble and D.F. Walls.  1987 {\it J. Opt. Soc. Am. B} {\bf 4} 10.
\bibitem{mandel} Ou Z Y, Hong C K and Mandel L 1987 {\it J. Opt. Soc. Am. B} {\bf 4} 1574
\bibitem{zoller}Walls D F and Z\"{o}ller P 1981 {\it Phys. Rev. Lett.} {\bf 47} 709 \\
Mandel L 1982 {\it Phys. Rev. Lett.} {\bf 49} 136 \\ Zhou P and Swain S 1999 {\it Phys. Rev. A}
{\bf 59} 3745 \\ Ficek Z and Swain S 1997 {\it J. Opt. Soc. Am. B} {\bf 14} 258
\bibitem{collet} Collet M J, Walls D F and Z\"{o}ller P 1984 {\it Opt. Commun.} {\bf 52} 145
\bibitem{swain} Zhou P and Swain S 1999 {\it Phys. Rev. A} {\bf 59} 841 \\ Zhou P and Swain S 1999
{\it Phys. Rev. A} {\bf 59} 1603
\bibitem{oliveira} de Oliveira F A M, Dalton B J and Knight P L 1987 {\it J. Opt. Soc. Am. B} {\bf 4}
1558
\bibitem{vogel} Ficek Z, Dalton B J and Knight P L 1994 {\it Phys. Rev. A} {\bf 50} 2594 \\
Vogel W and Blatt R 1992 {\it Phys. Rev. A} {\bf 45} 3319
\bibitem{vee} Lakshmi P A and Agarwal G S 1985 {\it Phys. Rev. A} {\bf 32} 1643
\bibitem{ladder} Ficek Z, Dalton B J and Knight P L 1995 {\it Phys. Rev. A} {\bf 51} 4062
\bibitem{dalton} Dalton B J, Ficek Z and Knight P L 1994 {\it Phys. Rev. A} {\bf 50} 2646
\bibitem{vogel2} Gr\"{u}nwald P and Vogel W 2012 {\it Phys. Rev. Lett.} {\bf 109} 013601 \\
Gr\"{u}nwald P and Vogel W 2013 {\it Phys. Rev. A} {\bf 88} 023837
\bibitem{gao} Gao S Y, Li F L and Zhu S Y 2005 {\it Phys. Lett. A} {\bf 335} 110 \\
Gao S Y, Li F L and Cai D L 2007 {\it J. Phys. B: At. Mol. Opt. Phys.} {\bf 40} 3893
\bibitem{vic} Menon S and Agarwal G S 1998 {\it Phys. Rev. A} {\bf 57} 4014 \\
Paspalakis E and Knight P L 1998 {\it Phys. Rev. Lett} {\bf 81} 293\\
Macovei M, Evers J and Keitel C H 2003, {\it Phys. Rev. Lett} {\bf 91} 233601
\bibitem{trap} Kosachiov D V, Matisov B G and Rozhdestvensky Y V 1992
{\it J. Phys. B: At. Mol. Opt. Phys.} {\bf 25} 2473
\bibitem{ionize} Li G X and Peng J S 1996 {\it Phys. Lett. A} {\bf 218} 49 \\
Paspalakis E, Patel A, Protopapas and Knight P L 1998 {\it J. Phys. B: At. Mol. Opt. Phys.} 
{\bf 31} 761
\bibitem{hemmer} Shahriar M S and Hemmer P R 1990 {\it Phys. Rev. Lett.} {\bf 65} 1865
\bibitem{probes} Yamamota K, Ichimura K and Gemma N 1998 {\it Phys. Rev. A} {\bf 58} 2460
\bibitem{eit1} Korsunsky E A, Leinfellner N, Huss A, Baluschev S and Windholz L 1999
{\it Phys. Rev. A} {\bf 59} 2302
\bibitem{eit2} Wilson E A, Manson N B, Wei C and Yang L 2005 {\it Phys. Rev. A} {\bf 72}
063813 \\
Wilson E A, Manson N B and Wei C 2005 {\it Phys. Rev. A} {\bf 72} 063814
\bibitem{scully} Li H, Sautenkov V A, Rostovtsev Y V, Welch G R, Hemmer P R and Scully M O
2009 {\it Phys. Rev. A} {\bf 80} 023820
\bibitem{sun} Sun H, Guo H, Bai Y F, Han D A, Fan S L and Chen X Z 2005 {\it Phys. Lett. A} {\bf 335}
68 \\
Agarwal G S, Dey T N and Menon S 2001 {\it Phys. Rev. A} {\bf 64} 053809
\bibitem{quench} Xu Q, Hu X M and Yin J W 2008 {\it J. Phys. B: At. Mol. Opt. Phys.} {\bf 41}
035503
\bibitem{dark} Luo B, Tang H and Guo H 2009 {\it J. Phys. B: At. Mol. Opt. Phys.} {\bf 42} 235505
\bibitem{sref} Gao S Y, Li F L and Zhu S Y 2002 {\it Phys. Rev. A} {\bf 66} 043806 \\
Gonzalo I, Ant\'{o}n M A, Carre\~{n}o F and Calder\'{o}n O G 2005 {\it Phys. Rev. A}
{\bf 72} 033809
\bibitem{note1} In all figures, we assume $\Delta_3 = \Delta_1 - \Delta_2$ to satisfy the condition
$\Delta_4 = 0$.
\bibitem{note2} For $\Omega_3 = 0$, the spectral squeezing is absent for all values of Rabi frequencies
and detunings provided the parameters satisfy the condition $\Omega_1 = \Omega_2$ and $\Delta_1 = -\Delta_2$.
\bibitem{note3} For the parameters $(\Phi = 0, \pi)$ considered in this paper, the expansion coefficients
$(a_{1i},a_{2i},a_{3i})$ are real. So, the normalization condition implies $a_{1i}^2 + a_{2i}^2 + a_{3i}^2 = 1$.
\end{thebibliography}
\end{document}